\documentclass[a4paper, 12pt]{article}
\usepackage{epsfig,amssymb,euscript,xspace, color}
\usepackage{amsmath,jheppub,mathtools}
\def\bea{\begin{eqnarray}}
\def\eea{\end{eqnarray}}
\def\be{\begin{equation}}
\def\ee{\end{equation}}



\begin{document}

\title{
Holographic chiral induced W-gravities  }
\author{Rohan R Poojary and  Nemani V Suryanarayana}
\affiliation{Institute of Mathematical Sciences \\ C.I.T Campus, Taramani \\ Chennai 600113, India}
\emailAdd{ronp, nemani@imsc.res.in}

\abstract{
We study boundary conditions for 3-dimensional higher spin gravity that admit asymptotic symmetry algebras expected of 2-dimensional induced higher spin theories in the light cone gauge. For the higher spin theory based on $sl(3, {\mathbb R}) \oplus sl(3, {\mathbb R})$ algebra, our boundary conditions give rise to one copy of classical $W_3$ and a copy of $sl(3,{\mathbb R})$ or $su(1,2)$ Kac-Moody symmetry algebra. We propose that the higher spin theories with these boundary conditions describe appropriate chiral induced W-gravity theories on the boundary. We also consider boundary conditions of spin-3 higher spin gravity that admit $u(1) \oplus u(1)$ current algebra.}

\maketitle



\section{Introduction}

Long ago in 1987 Polyakov in a seminal paper \cite{Polyakov:1987zb} addressed the problem of induced gravity \cite{Polyakov:1981rd} in two dimensions in light cone gauge, referred to as the chiral induced gravity (CIG) theory. This enabled uncovering of an $sl(2, {\mathbb R})$ current algebra worth of symmetry which in turn lead to the determination of all correlation functions in that theory. 

Following Polyakov's work and the discovery of W-symmetries (see \cite{Bouwknegt:1992wg} for a review) people studied the induced W-gravity theories in a particular light-cone gauge.\footnote{See, for instance, \cite{Ooguri:1991by} and references therein.} These theories are dubbed the chiral induced W-gravity theories (CIWG). They are expected to have $sl(n, {\mathbb R})$-type current algebra symmetries (generalizing the $sl(2, {\mathbb R})$ current algebra symmetry of the CIG of Polyakov), and local actions. See, for instance \cite{Bershadsky:1989mf, Ooguri:1991by}.

The CIWG theories provide some important examples of 2-dimensional field theories that admit only one copy of Virasoro algebra. Some recent examples of such theories have  been defined by \cite{Compere:2013aya, Compere:2013bya, Avery:2013dja, Apolo:2014tua} via holography. It is an interesting question to study if CIWG theories also admit holographic descriptions. 

In an earlier paper \cite{Avery:2013dja} (see also \cite{Apolo:2014tua}) we proposed a set of boundary conditions for $AdS_3$ gravity which admitted the same symmetry algebra as Polyakov's 2d chiral induced gravity \cite{Polyakov:1987zb}. These were further generalized and studied in \cite{Apolo:2014tua} which provided evidence for the proposed duality. We will refer to the boundary conditions of \cite{Avery:2013dja, Apolo:2014tua} as Chiral Induced Gravity (CIG) boundary conditions. The asympototic symmetry algebra of the CIG boundary conditions were a copy of a Virasoro and a copy of $sl(2, {\mathbb R})$ Kac-Moody algebra with level $k$ given by $c/6$. In this paper we generalise these results towards studying chiral induced W-gravities (CIWG) holographically. We argue that the bulk theory should be a higher spin theory with one higher spin gauge field corresponding to each higher spin field in the induced W-gravity theory. We, therefore, propose that the 3d higher spin theories based on $sl(n, {\mathbb R})$ Chern-Simons theories \cite{Blencowe:1988gj} admit boundary conditions that are good candidates to describe a suitable chiral induced W-gravity with spins from 2 to $n$. 

Towards this, we provide and study a set of boundary conditions for the case of $n=3$ and compute their asymptotic symmetry algebras. We find that, in this case, the higher spin theory with our boundary conditions admit one copy of classical $W_3$ algebra and an $sl(3, {\mathbb R})$ (or an $su(1,2)$) current algebra. As a byproduct we also provide a generalization of the boundary conditions of \cite{Compere:2013bya} to this higher spin theory and compute the corresponding symmetry algebra.

The rest of the paper is organised as follows. In section 2 we briefly review the results of \cite{Avery:2013dja, Apolo:2014tua} and translate them into the $sl(2, {\mathbb R}) \oplus  sl(2, {\mathbb R}) $ Chern-Simons formalism of $AdS_3$ gravity. In section 3 we generalise the results of section 2 to higher spin theory with spin 2 and spin 3 fields in the $sl(3, {\mathbb R}) \oplus  sl(3, {\mathbb R}) $ Chern-Simons formalism. In section 4 we compute the asymptotic symmetry algebras. In the last section we provide some comments, discuss open issues  and conclude. 
 
\section{CIG boundary conditions}
In an earlier paper \cite{Avery:2013dja} with S. Avery we proposed a set of boundary conditions for $AdS_3$ gravity (in the metric formalism)
\bea
\label{ads3action}
S_{d=3} &=& - \frac{1}{16 \pi G} \int_{\cal M} d^3x \, \sqrt{|g|} \, \left(R + \frac{2}{l^2}\right)  \cr 
&& ~~~~~~ - \frac{1}{8\pi G} \int_{\partial\cal M} d^2x \, \sqrt{|h|} \, K + \frac{1}{8\pi G} \int_{\partial\cal M} d^2x \, \frac{1}{l} \sqrt{|h|} + \cdots
\eea
that admitted an $sl(2, {\mathbb R})$ current algebra as their asymptotic symmetry algebra. The motivation was to provide a holographic description of the induced gravity studied in a light cone gauge by Polyakov \cite{Polyakov:1987zb}. Here we begin with a brief review of those results with clarifications and generalizations. Let us consider the following boundary conditions \cite{Avery:2013dja, Apolo:2014tua} for the metric

\begin{equation}
\label{apsbdyconds}
\begin{aligned}
g_{rr} &= \frac{l^2}{r^2} + {\cal O}(r^{-4}), ~~ g_{r+} = {\cal O}(r^{-1}), ~~ g_{r-} = {\cal O}(r^{-3}), \\
g_{+-} &= - \frac{r^2}{2} + {\cal O}(r^0), ~~ g_{--} =  {\cal O}(r^0), \\
g_{++} &= r^2 F(x^+, x^-) + {\cal O} (r^0) ,
\end{aligned}
\end{equation}
where $x^+, x^-$ are treated to be the boundary coordinates and $r$ is the radial coordinate with the asymptotic boundary at $r^{-1} = 0$. One can write a general non-linear solution of $AdS_3$ gravity in Fefferman--Graham
coordinates  \cite{Skenderis:1999nb} as:
\begin{equation}
\label{nlsoln1}
ds^2 = l^2 \frac{dr^2}{r^2} + r^2 \left[ g^{(0)}_{ab} + \frac{l^2}{r^2} \, g^{(2)}_{ab} + \frac{l^4}{r^4} g^{(4)}_{ab} \right] dx^a dx^b.
\end{equation}
Therefore, the full set of non-linear solutions consistent with our boundary conditions is obtained when
\begin{equation}
\label{nlsoln2}
\begin{aligned}
g^{(0)} _{++ } &= F(x^+, x^-), ~~ g^{(0)}_{+-} = -\frac{1}{2}, ~~ g^{(0)}_{--} = 0, \\
g^{(2)}_{++} &= \kappa(x^+, x^-), ~~  g^{(2)}_{+-} = \sigma(x^+,x^-), 
      ~~ g^{(2)}_{--} =\tilde \kappa (x^+, x^-),\\
g^{(4)}_{ab} &= \frac{1}{4} g^{(2)}_{ac} g_{(0)}^{cd} g^{(2)}_{db} \, ,
\end{aligned}\end{equation}
where in the last line $g_{(0)}^{cd}$ is $g^{(0)}_{cd}$ inverse. Imposing the equations of motion $R_{\mu\nu} - \frac{1}{2} R \, g_{\mu\nu} - \frac{1}{l^2} g_{\mu\nu}=0$ one finds that these equations are satisfied for $\mu, \nu = +, -$. Then the remaining three equations coming from $(\mu, \nu)  = (r,r), (r,+), (r,-)$ impose the following relations:
\bea
\label{nlsoln3}
\sigma (x^+, x^-) &=& \frac{1}{2} [\partial_-^2 F - 2 \, \tilde \kappa  \, F] \cr
\kappa (x^+, x^-) &=& \kappa_0 (x^+) + \frac{1}{2} [ \partial_+ \partial_- F+ 2 \, \tilde \kappa \, F^2 - F \, \partial_-^2 F  - \tfrac{1}{2} (\partial_-F)^2]
\eea
and
\bea
\label{virasorowardidentity}
2 \, (\partial_+ + 2 \, \partial_- F + F \, \partial_-) \, \tilde \kappa = \partial_-^3 F \, .
\eea
This last equation may be recognised as the Virasoro Ward identity of Polyakov\cite{Polyakov:1987zb} expected from the 2d CIG. This Ward identity is integrable. To find the solution, inspired by Polyakov, let us parametrize $F$ = $-\frac{\partial_+ f}{\partial_- f}$. With this parametrization one can show that the above constraint (\ref{virasorowardidentity}) can be cast into the following form:
\bea
(\partial_-f \, \partial_+ - \partial_+f \, \partial_-) \Big[ 4\, (\partial_-f)^{-2} \, \tilde \kappa - (\partial_-f)^{-4} \, [ 3 \, (\partial_-^2f)^2 - 2 \, \partial_-f (\partial_-^3f)]\Big] = 0
\eea
For an arbitrary $f(x^+, x^-)$ the general solution to this equation is
\bea
\label{nlsoln4}
\tilde\kappa(x^+, x^-) = \frac{1}{4}G[f] (\partial_-f)^{2}+\frac{1}{4} (\partial_-f)^{-2} \, [ 3 \, (\partial_-^2f)^2 - 2 \, \partial_-f (\partial_-^3f)] 
\eea
where $G[f]$ is an arbitrary functional of $f(x^+,x^-)$. The second term in the solution may be recognized as the Schwarzian derivative of $f$ with respect to $x^-$.  

Along with this solution (\ref{nlsoln4}) for $\tilde \kappa$ the configurations in (\ref{nlsoln1} - \ref{nlsoln3}) provide the most general solutions consistent with the boundary conditions in (\ref{apsbdyconds}). 

The $AdS_3$ gravity with the boundary conditions (\ref{apsbdyconds}) should provide a  holographic description of the 2d CIG with $F$ playing the role of its dynamical field. However, the classical solutions of the 2d CIG should correspond to bulk solutions with $\tilde \kappa$ either vanishing or an appropriate non-zero constant.  In the latter case one needs to add additional boundary terms to the action (\ref{ads3action}), see \cite{Compere:2013bya}, \cite{Avery:2013dja}. When $\tilde \kappa =0$ one gets the solutions appropriate to asymptotically Poincare $AdS_3$, where as $\tilde \kappa = - 1/4$ corresponds to the solutions considered in \cite{Avery:2013dja}.\footnote{Let us note that these solutions also include those of \cite{Compere:2013bya} when one takes $\tilde \kappa = \Delta$ and $\partial_- F =0$.} 

The asymptotic symmetries of configurations in (\ref{nlsoln1} - \ref{nlsoln3}) are generated by the following vector field:
\bea
\xi &=& -\frac{1}{2} [ \lambda_\kappa'(x^+) + \partial_- \lambda (x^+, x^-) + {\cal O} (r^{-4}) ] \, r \, \partial_r  + [\lambda_\kappa(x^+) + \frac{l^2}{2r^2} \partial_-^2 \lambda (x^+, x^-) + {\cal O} (r^{-4})] \, \partial_+ \cr
&& + [ \lambda(x^+, x^-)+ \frac{l^2}{2r^2} (\lambda_\kappa''(x^+) + 2 \, F \, \partial_-^2 \lambda (x^+, x^-)  + \partial_+ \partial_- \lambda(x^+, x^-) ) + {\cal O} (r^{-4})]  \, \partial_- \, . 
\eea
To compare the answers with \cite{Avery:2013dja} we consider the case of $\tilde \kappa = - 1/4$. Then the Virasoro Ward identity can be solved to get $F = f(x^+) + g(x^+) \, e^{i\, x^-} + \bar g(x^+) \, e^{-i \, x^-}$ which in turn implies 
\bea
\kappa = \kappa_0 (x^+) - \frac{1}{4} f^2 + \frac{1}{2} (e^{2i x^-} g^2 + e^{-2i x^-} \bar g^2) + \frac{i}{2} (e^{i x^-} g' - e^{-i x^-} \bar g') \, . 
\eea
At the same time keeping $\tilde \kappa$ fixed under the asymptotic diffeomorphisms requires that 
\bea
\lambda (x^+, x^-)  = \lambda_f (x^+) + e^{i \, x^-} \,\lambda_g (x^+) + e^{-i \, x^-} \, \bar \lambda_{\bar g} (x^+)
\eea
These induce the following transformations on $f(x^+)$, $g(x^+)$, $\bar g(x^+)$ and $\kappa_0(x^+)$:
\bea
\label{apsvariations}
\delta_\xi f(x^+) &=&  \lambda_f'  +2i \, (\bar g \, \lambda_g - \bar \lambda_{\bar g} \, g) + (f \lambda_\kappa)', ~~ \delta_\xi g = \lambda_g'  + i \, (f \, \lambda_g - g \, \lambda_f ) + (g \lambda_\kappa)', \cr
\delta_\xi \bar g &=& \bar \lambda_{\bar g}'  - i \, (f \, \bar \lambda_{\bar g} - \bar g \, \lambda_f ) + (\bar g \lambda_\kappa)', ~~
\delta_\xi \kappa_0 = \lambda_\kappa \, \kappa_0' + 2 \, \kappa_0 \, \partial_+ \lambda_\kappa -\frac{1}{2} \partial_+^3 \lambda_\kappa 
\eea
These expressions can be summarized as
\bea
\delta_\xi J^a &=& \partial_+ \lambda^a - i \, {f^a}_{bc} J^b \, \lambda^c + \partial_+ (J^a \, \lambda) \cr
&=& \partial_+ (\lambda^a + J^a \, \lambda) - i  \, {f^a}_{bc} J^b \, (\lambda^c + J^c \, \lambda)
\eea
where $\{J^{(-1)}, J^{(0)}, J^{(1)}\} = \{\bar g, f, g \}$ and $\{\lambda^{(-1)}, \lambda^{(0)}, \lambda^{(1)}\} = \{\bar \lambda_{\bar g} , \lambda_f, \lambda_g\}$ and $\lambda = \lambda_\kappa$.

The charges for these symmetries were shown to be integrable and finite, and computed in \cite{Avery:2013dja}. In the next section we would like to recover these results from the first order formalism of $AdS_3$ gravity as a warm up for generalisation to the higher spin case.
\subsection{Holographic CIG in first order formalism}
It is well-known \cite{Achucarro:1987vz,Witten:1988hc} that the $AdS_3$ gravity in the Hilbert-Palatini formulation can be recast as a Chern-Simons gauge theory with action\footnote{As is standard the symbol $tr$ is understood to be $\tfrac{1}{2 {\rm Tr} L_0^2} {\rm Tr}$ where ${\rm Tr}$ is the ordinary matrix trace.}
\bea
S[A, \tilde A] = \frac{k}{4\pi} \int {\rm tr} (A\wedge A + \frac{2}{3} A \wedge A \wedge A) - \frac{k}{4\pi} \int {\rm tr} (\tilde A \wedge \tilde A + \frac{2}{3} \tilde A \wedge \tilde A \wedge \tilde A) 
\eea
up to boundary terms, where the gauge group is $SL(2, {\mathbb R})$. These are related to veilbein and spin connection through $A = \omega^a + \frac{1}{l} e^a$ and $\tilde A = \omega^a - \frac{1}{l} e^a$. The equations of motion are $F = dA + A \wedge A = 0$ and $\tilde F = d\tilde A + \tilde A \wedge \tilde A = 0$. See appendix A for details on the most general solutions to these flatness conditions. 

Next we will write the solutions of $AdS_3$ gravity consistent with (\ref{apsbdyconds})   in CS language. For this we simply specialize the flat connections given in appendix A to
\bea
A &=& b^{-1} \partial_r b \, dr + b^{-1} [(L_1 + a^{(-)}_+ \,  L_{-1}+  a^{(0)}_+ \, L_0) \,  dx^+ + (a^{(-)}_- \, L_{-1}) \, dx^-] \, b \,\cr
\tilde A &=& b \, \partial_r b^{-1}  \, dr + b \,  [(\tilde a^{(0)}_+ \, L_0 + \tilde a^{(+)}_+ \, L_1 + \tilde a^{(-)}_+ \, L_{-1}) \, dx^+ + ( \tilde a^{(+)}_- \, L_1- L_{-1} ) \, dx^- ]\, b^{-1}
\eea
where $b = e^{L_0 \, \ln \frac{r}{l}}$ and all the functions are taken to depend on both the boundary coordinates $(x^+, x^-)$. The equations of motion impose the following conditions:
\bea
\label{sl2rsolns}
a^{(-)}_-  &=& \frac{1}{2} \partial_- a^{(0)}_+, ~~ a^{(-)}_+ = -\kappa_0 (x^+) + \frac{1}{4} (a^{(0)}_+)^2 + \frac{1}{2} \partial_+ a^{(0)}_+ \cr
\tilde a^{(0)}_+ & =& - \partial_- \tilde a^{(-)}_+,  ~~
\tilde a^{(+)}_+ = - \tilde a^{(-)}_+ \, \tilde a^{(+)}_- - \frac{1}{2} \partial_- \tilde a^{(0)}_+, 
\eea
and 
\bea
(\partial_+ + 2 \, \partial_- \tilde a^{(-)}_+ + \tilde a^{(-)}_+ \, \partial_-) \, \tilde a^{(+)}_- = \frac{1}{2} \partial_-^3 \tilde a^{(-)}_+ 
\eea
The last equation is a Virasoro Ward identity and it can be solved as before. To obtain metric in the FG gauge we need to impose $a^{(0)}_+ = \tilde a^{(0)}_+$. With this choice it is easy to see that the metric obtained matches exactly with the solution given above in (\ref{nlsoln1} - \ref{nlsoln3}) with $F = \tilde a^{(-)}_+$ and $\tilde \kappa = \tilde a^{(+)}_-$. 

To be able to define a variational principle that admits a fluctuating $F$, we add the following boundary action:
\bea
S_{bdy.} = -\frac{k}{4\pi} \int d^2x \,  tr (L_0 \, [a_+, a_-]) - \frac{k}{4\pi} \int d^2x \,  tr (L_0 \, [\tilde a_+, \tilde a_-] -2 \, \tilde \kappa_0 \, L_1 \, \tilde a_+) 
\eea
where $\tilde \kappa_0$ is some constant. Then it is easy to see that the variation of the full action gives
\bea
\delta S_{total} = \frac{k}{2\pi} \int d^2x \, (\tilde \kappa - \tilde \kappa_0) \, \delta F.
\eea
In showing this we have to use all the constraints in (\ref{sl2rsolns}) coming from the equations of motion except the Virasoro Ward identity. Therefore, we again have two ways to impose the variational principle $\delta S = 0$: (i) $\delta F = 0$ and (ii) $\tilde \kappa = \tilde \kappa_0$. The former is the usual Brown-Henneaux \cite{Brown:1986nw} type Dirichlet boundary condition. We therefore consider the latter.
\subsection{Residual gauge transformations}
To analyze the asymptotic symmetries in the CS language we seek the residual gauge transformations that leave $\tilde \kappa$ fixed, and the above flat connections form-invariant.

The gauge transformations act as $\delta_{\Lambda}  A = d \Lambda + [A,  \Lambda]$ which in turn act as $\delta_{\lambda} a =  d \lambda + [a, \Lambda]$ where $ A = b^{-1} \, a \, b + b^{-1} \, d b$ with $\Lambda = b^{-1} \, \lambda \, b$ (and similarly on the right sector gauge field $\tilde a$ with parameter labeled $\tilde \lambda$). The resulting gauge parameters are
\bea
\lambda &=& \lambda^{(-)} (x^+, x^-) \, L_{-1} + [a^{(0)}_+ \, \lambda^{(+)} (x^+) - \partial_+ \lambda^{(+)} (x^+)] \, L_0 + \lambda^{(+)}(x^+) \, L_{1}  \cr
 \tilde \lambda &=& \tilde \lambda^{(-)} \, L_{-1}  - \partial_- \tilde \lambda^{(-)} \, L_0 - [\tilde a^{(+)}_- \, \tilde \lambda^{(-)} -  \frac{1}{2} \partial_-^2 \bar \lambda^{(-)}] \, L_1, 
\eea
that induce the variations
\bea
\delta_\lambda a^{(0)}_+ &=& 2 \, [\lambda^{(-)} -  a^{(-)}_+ \, \lambda^{(+)}] - \partial_+ [\partial_+ \lambda^{(+)} - a^{(0)}_+ \,\lambda^{(+)} ] \cr
\delta_\lambda a^{(-)}_+ &=& \partial_+ \lambda^{(-)} +  a^{(0)}_+ \, \lambda^{(-)} +  a^{(-)}_+  \, [ \partial_+ \lambda^{(+)} - a^{(0)}_+ \lambda^{(+)} \,  ]
\eea
and
\bea
\delta_{\tilde \lambda} \tilde a^{(-)}_+ &=& (\partial_+  +  \tilde a^{(-)}_+ \, \partial_-   - \partial_- \tilde a^{(-)}_+ ) \, \tilde \lambda^{(-)}  \cr
\delta_{\tilde \lambda} \tilde a^{(+)}_- &=& - \tilde \lambda^{(-)} \, \partial_- \tilde a^{(+)}_- - 2 \, \tilde a^{(+)}_- \, \partial_- \tilde \lambda^{(-)} + \frac{1}{2} \, \partial_-^3 \tilde \lambda^{(-)}
\eea
respectively. In the global case when we hold $\tilde a^{(+)}_-$ fixed at $-1/4$ we find that $\tilde \lambda^{(-)} = \lambda_f + e^{i\, x^-} \lambda_g + e^{-i \, x^-} \, \bar \lambda_{\bar g}$. When we make a gauge transformation to ensure that we remain the FG coordinates for the metric we need to impose  
\bea 
(\delta_\lambda a^{(0)}_+ - \delta_{\tilde \lambda} \tilde a^{(0)}_+)\Big{|}_{a^{(0)}_+=\tilde a^{(0)}_+} = 0
\eea
We find that this condition drastically reduces the number of independent residual gauge parameters down to four functions of $x^+$. In particular, the function $\lambda^{(-)} (x^+, x^-)$ is determined to be
\bea
&& \lambda^{(-)} (x^+, x^-) = -\frac{1}{4} \lambda^{(+)} (\bar g \, e^{-i \, x^-} - g \, e^{i \, x^-})^2  - \, \kappa_0 \, \lambda^{(+)} + \frac{1}{2} \partial_+^2 \lambda^{(+)} + \frac{i}{2} (g \, e^{i \, x^-} - \bar g \, e^{-i \, x^-})  \partial_+ \lambda^{(+)}\cr
&& + \frac{1}{2} [ (\lambda_g \, e^{i \, x^-} + \bar \lambda_{\bar g} \, e^{-i \, x^-}) \, f - (g \, e^{i \, x^-} + \bar g \, e^{-i \, x^-}) \, \lambda_f 
 - i (\partial_+ \lambda_g \, e^{i\, x^-} - \partial_+ \bar\lambda_{\bar g} \, e^{-i\, x^-})]
\eea
These induce the following transformations:
\bea
\delta_\lambda f &=& \lambda_f' +2i \, (\bar g \, \lambda_g - \bar \lambda_{\bar g} \, g), ~~ \delta_\lambda g = \lambda_g' + i \, (f \, \lambda_g - g \, \lambda_f ), ~~ \delta_\lambda \bar g = \bar \lambda_{\bar g}' - i \, (f \, \bar \lambda_{\bar g} - \bar g \, \lambda_f ), \cr
\delta \kappa_0 &=& \lambda^{(+)} \, \kappa_0' + 2 \kappa_0 \, \partial_+ \lambda^{(+)} -\frac{1}{2} \partial_+^3 \lambda^{(+)}
\eea
We could have obtained this result starting with the left sector $a$ to be $a = [L_1 - \kappa_0 (x^+) \, L_{-1}] \, dx^+$. 

Now, comparing this result with (\ref{apsvariations}) one finds that $\{\lambda_f, \lambda_g, \bar \lambda_{\bar g}, \lambda^{(+)}\}$ are not quite the parameters in (\ref{apsvariations}) that correspond to the asymptotic symmetry vector fields of \cite{Avery:2013dja}. For this it turns out that we have to redefine the gauge parameters 
\bea
\{\lambda_f, \lambda_g, \bar \lambda_{\bar g}, \lambda\} \rightarrow \{\lambda_f + f \, \lambda, \lambda_g + g \, \lambda, \bar \lambda_{\bar g} + \bar g \, \lambda \} 
\eea
Then the transformations read
\bea
\delta_{\tilde \lambda} f &=& \lambda_f'  +2i \, (\bar g \, \lambda_g - \bar \lambda_{\bar g} \, g) + (f \lambda^{(+)})', ~~ \delta_{\tilde \lambda} g = \lambda_g'  + i \, (f \, \lambda_g - g \, \lambda_f ) + (g \lambda^{(+)})', \cr
\delta_{\tilde \lambda} \bar g &=& \bar \lambda_{\bar g}'  - i \, (f \, \bar \lambda_{\bar g} - \bar g \, \lambda_f ) + (\bar g \lambda^{(+)})', ~~
\delta \kappa_0 = \lambda^{(+)} \, \kappa_0' + 2 \kappa_0 \, \partial_+ \lambda^{(+)} -\frac{1}{2} \partial_+^3 \lambda^{(+)}
\eea
which match exactly with those in (\ref{apsvariations}). If we compute the charges and the algebra of these symmetries it can be seen that they match exactly with those of \cite{Avery:2013dja}. In the next section we turn to generalization to the higher spin theories. 

\section{Chiral boundary conditions for $SL(3,{\mathbb R})$ higher spin theory}

In this section we are interested in proposing boundary conditions for higher spin theories such that they holographically describe  appropriate chiral induced $W$-gravity theories. To demonstrate the result we restrict to the simplest higher spin theory in three dimensions that contains a spin-2 field and a spin-3 field. In the first order formalism the theory is formulated on the same lines as $AdS_3$ gravity but with the gauge group replaced by $SL(3,{\mathbb R})$ (or $SU(1,2)$) \cite{Blencowe:1988gj,Campoleoni:2011hg}. The dirichlet boundary conditions of this theory were considered by Campoleoni {\it et} {\it al} \cite{Campoleoni:2011hg} and they showed that the asymptotic symmetry algebra is two commuting copies of classical $W_3$ algebra with central charges. 

We now turn to generalising the boundary conditions of the section (2.1) to the 3-dimensional higher spin theory based on two copies of $sl(3, {\mathbb R})$ or $su(1,2)$ algebra \cite{Blencowe:1988gj,Campoleoni:2011hg}. Motivated by the CIG boundary conditions of the previous section we write the connections again as deformations of $AdS_3$ solution. We work in the principal embedding basis for the gauge algebra. Our conventions very closely follow those of \cite{Campoleoni:2011hg} and may be found in appendix B. We start with the following ansatz for the gauge connections:
\bea
\label{theansatz}
A &=& b^{-1} \partial_r b \, dr + b^{-1} \left[ \left(L_1 - \kappa \,  L_{-1} - \omega \, W_{-2} \right) \, dx^+ \right] \, b  \\
\tilde A &=& b \, \partial_r b^{-1}  \, dr + b \, \left[ \left( - L_{-1} + \tilde \kappa \, L_1 + \tilde \omega \, W_2 \right)  \, dx^-
+ \left(\sum_{a = -1}^1 f^{(a)} L_a + \sum_{i=-2}^2 g^{(i)} W_i \right) \, dx^+ \right] b^{-1} \,  \nonumber 
\eea
where $b$ is again $e^{L_0 \, \ln \frac{r}{l}}$. Note that, as in $sl(2, {\mathbb R})$ case, our ansatz is the Dirichlet one of \cite{Campoleoni:2011hg} for the left sector. Similarly, the right sector includes the right sector ansatz of previous section as a special case.\footnote{A similar ansatz has been considered recently in \cite{deBoer:2014fra}. }  All the coefficients of the algebra generators above are understood to be functions of $x^+$ and $x^-$.

Imposing flatness conditions on $A$ and $\tilde A$ demand
\bea
\partial_- \kappa &=& 0, ~~ \partial_- \omega = 0, 
\eea
and 
\newpage
\bea
\partial_- f^{(-1)} + f^{(0)} &=& 0, \cr
\partial_- f^{(0)} + 2 \, f^{(1)} +2 \, \tilde \kappa \, f^{(-1)} - 16 \, \alpha^2 \, \tilde \omega \, g^{(-2)} &=& 0, \cr
-\partial_+ \tilde \kappa + \partial_- f^{(1)} + \tilde \kappa \, f^{(0)} - 4 \, \alpha^2 \, \tilde \omega \, g^{(-1)}&=& 0 \cr
&& \cr
\partial_- g^{(-2)} + g^{(-1)} &=& 0, \cr
\partial_- g^{(-1)} + 2 \, g^{(0)} + 4 \, \tilde \kappa \, g^{(-2)} &=& 0, \cr
\partial_- g^{(0)} + 3 \, g^{(1)} + 3 \, \tilde \kappa \, g^{(-1)} &=& 0, \cr
\partial_- g^{(1)} + 4 \, g^{(2)} + 2 \, \tilde \kappa \, g^{(0)} + 4 \, \tilde \omega \, f^{(-1)} &=& 0, \cr
-\partial_+ \tilde \omega + \partial_- g^{(2)} + \tilde \kappa \, g^{(1)} + 2 \, \tilde \omega \, f^{(0)} &=& 0.
\eea
These equations enable one to solve for \{$f^{(0)}$, $f^{(1)}$, $g^{(-1)}$, $g^{(0)}$, $g^{(1)}$, $g^{(2)}$\} in terms of \{$\tilde \kappa$, $\tilde \omega$, $f^{(-1)}$, $g^{(-2)} \}$ and their derivatives, provided the functions  \{$\tilde \kappa$, $\tilde \omega$, $f^{(-1)}$, $g^{(-2)} \}$ satisfy the constraints coming from  the 3rd and the 8th equations:
\bea
\label{w3ward1}
(\partial_+ + 2 \, \partial_-f^{(-1)} + f^{(-1)} \, \partial_-) \, \tilde \kappa -  \alpha^2 \, (12 \, \partial_- g^{(-2)} + 8 \, g^{(-2)} \partial_-) \, \tilde \omega  = \frac{1}{2} \partial_-^3 f^{(-1)},
\eea
\bea
\label{w3ward2}
&& 12 \, (\partial_+ + 3 \, \partial_- f^{(-1)} + f^{(-1)} \partial_-) \, \tilde \omega + (10 \, \partial_-^3 g^{(-2)} + 15 \, \partial_-^2 g^{(-2)} \, \partial_- \!+ 9 \, \partial_- g^{(-2)} \partial_-^2 \!+ 2 \, g^{(-2)} \partial_-^3) \, \tilde \kappa \cr
&& ~~~~~~~~~~~~~~~~~~~  - 16 \, (2 \, \partial_- g^{(-2)} \!+ g^{(-2)} \, \partial_-) \, \tilde \kappa^2 = \frac{1}{2} \partial_-^5 g^{(-2)}.
\eea
 We point out that these are the Ward identities that the induced $W_3$ gravity theory is expected to satisfy. See Ooguri {\it et} {\it al} \cite{Ooguri:1991by} for a comparison. These have also appeared recently in \cite{deBoer:2014fra} in a related context.
 
Next, we seek the residual gauge transformations of our solutions. Defining the gauge parameter to be $\lambda = \lambda^{(a)} L_a + \eta^{(i)} \, W_i$ and imposing the conditions that the gauge field configuration $a= (L_1 - \kappa \,  L_{-1} - \omega \, W_{-2}) \, dx^+$ is left form-invariant leads to the following conditions \cite{Campoleoni:2011hg}:
\bea
\partial_- \lambda^{(a)} = \partial_- \eta^{(i)}  &=& 0, \cr
\partial_+ \lambda^{(0)} + 2 \, \lambda^{(-1)} + 2 \, \kappa \, \lambda^{(1)} - 16 \, \alpha^2 \,  \omega \, \eta^{(2)} &=& 0, \cr
\partial_+ \lambda^{(1)} + \lambda^{(0)} &=& 0,  \cr
\partial_+ \eta^{(-1)} + 4\, \eta^{(-2)} + 2 \, \kappa \, \eta^{(0)} + 4 \, \omega \, \lambda^{(1)} &=& 0, \cr
\partial_+ \eta^{(0)} + 3 \, \eta^{(-1)}  + 3 \, \kappa \, \eta^{(1)} &=& 0, \cr
\partial_+ \eta^{(1)} + 2 \, \eta^{(0)}  + 4 \, \kappa \, \eta^{(2)} &=& 0, \cr
\partial_+ \eta^{(2)} + \eta^{(1)} &=& 0.
\eea
Under these (relabelling $\lambda^{(1)} \rightarrow \lambda$ and $\eta^{(2)} \rightarrow \eta$) we have
\bea
\label{deltakappa}
\delta \kappa = \lambda \, \kappa' + 2 \, \lambda' \, \kappa - \frac{1}{2} \lambda''' - 8\, \alpha^2  \, \eta \, \omega' - 12 \, \alpha^2 \, \omega \, \eta'
\eea
\bea
\label{deltaomega}
\delta \omega = \lambda \, \omega' + 3 \, \omega \, \lambda'- \frac{8}{3} \, \kappa \, (\kappa \, \eta' + \eta \, \kappa') + \frac{1}{4} (5 \, \kappa' \, \eta'' + 3 \, \eta' \, \kappa'') + \frac{1}{6} \, (5 \, \kappa \, \eta''' + \eta \, \kappa''') - \frac{1}{24} \eta^{'''''} \nonumber \\
\eea
We parametrize the residual gauge transformations of $\tilde a$ by the gauge parameters $\tilde \lambda = \tilde \lambda^{(a)} L_a + \tilde \eta^{(i)} \, W_i$. The constraints on this parameter are
\bea
\tilde \lambda_0 + \partial_- \tilde \lambda^{(-1)} &=& 0, \cr
\partial_- \tilde \lambda_0+ 2 \, \tilde \lambda^{(1)} + 2\, \tilde \kappa \, \tilde \lambda^{(-1)} - 16\, \alpha^2 \, \tilde \omega \, \tilde \eta^{(-2)} &=& 0, \cr
\partial_-\tilde \eta^{(-2)} + \tilde \eta^{(-1)} &=&0, \cr
\partial_- \tilde \eta^{(-1)} + 2\, \tilde \eta^{(0)} + 4 \, \tilde \eta^{(-2)} \, \tilde \kappa &=& 0, \cr
\partial_- \tilde \eta^{(0)} + 3 \, \tilde \eta^{(1)} +3 \, \tilde \eta^{(-1)} \, \tilde \kappa &=& 0, \cr
\partial_- \tilde \eta^{(1)} + 4 \, \tilde \eta^{(2)} + 2 \, \tilde \eta^{(0)} \, \tilde \kappa + 4 \, \tilde \lambda^{(-1)} \, \tilde \omega &=& 0. 
\eea
These induce the following variations:
\bea
\delta \tilde \kappa = - 2 \, \tilde \kappa \, \partial_- \tilde \lambda^{(-1)} - \tilde \lambda^{(-1)} \partial_- \tilde \kappa + 8 \, \alpha^2 \, \tilde \eta^{(-2)} \, \partial_- \tilde \omega + 12 \, \alpha^2 \, \tilde \omega \, \partial_- \tilde \eta^{(-2)} + \frac{1}{2} \partial_-^3 \tilde\lambda^{(-1)} 
\eea
\bea
\delta \tilde \omega &=& - \tilde \lambda^{(-1)} \, \partial_- \tilde \omega - 3 \, \tilde \omega \, \partial_-\lambda^{(-1)} + \frac{8}{3} \tilde \kappa \, ( \tilde \kappa \, \partial_- \tilde \eta^{(-2)} + \tilde \eta^{(-2)} \, \partial_- \tilde \kappa)  \cr
&& ~~~ - \frac{1}{4} (5 \, \partial_-\tilde \kappa \, \partial_-^2 \tilde \eta^{(-2)} + 3 \, \partial_- \tilde \eta^{(-2)}\, \partial_-^2 \tilde \kappa) - \frac{1}{6} (5 \, \tilde \kappa \, \partial_-^3 \tilde \eta^{(-2)} + \tilde \eta^{(-2)} \, \partial_-^3 \tilde \kappa) + \frac{1}{24} \partial_-^5 \tilde \eta^{(-2)}  \nonumber \\
\eea
\bea
\delta f^{(-1)} &= & \partial_+ \tilde \lambda^{(-1)} + f^{(-1)} \, \partial_- \tilde \lambda^{(-1)} - \tilde \lambda^{(-1)} \, \partial_- f^{(-1)} + \frac{32}{3} \, \alpha^2 \, \tilde \kappa \,  ( g^{(-2)} \, \partial_- \tilde \eta^{(-2)} - \tilde \eta^{(-2)} \,  \partial_- g^{(-2)}) \cr
&& \!\! + \alpha^2 (\partial_- g^{(-2)} \, \partial_-^2 \tilde \eta^{(-2)} - \partial_- \tilde \eta^{(-2)} \, \partial_-^2  g^{(-2)}) - \frac{2}{3} \, \alpha^2 \, ( g^{(-2)} \, \partial_-^3 \tilde \eta^{(-2)} - \tilde \eta^{(-2)} \, \partial_-^3 g^{(-2)} )
\eea
\bea
\delta g^{(-2)} = \partial_+ \tilde \eta^{(-2)} + f^{(-1)} \, \partial_- \tilde \eta^{(-2)} - \tilde \lambda^{(-1)} \, \partial_- g^{(-2)} + 2 \, (g^{(-2)} \, \partial_- \tilde \lambda^{(-1)} - \tilde \eta^{(-2)} \, \partial_- f^{(-1)})
\eea
For the residual gauge transformations to be global symmetries of the boundary theory one needs to impose the variational principle $\delta S = 0$ as well. We add the following boundary action:
\bea
&& S_{bdy.} \cr
&& ~ = \frac{k}{4\pi} \int \, d^2x \, tr \,  (- L_0 [\tilde a_+, \tilde a_-] + 2 \, \tilde \kappa_0 \, L_1 \, \tilde a_+ + \frac{1}{2 \, \alpha} \, W_0 \{ \tilde a_+, \tilde a_- \} + \frac{1}{3} \tilde a_+ \, \tilde a_- + 2 \, \tilde \omega_0 \, W_2 \,\tilde a_+ ) \nonumber \\
\eea
With this the variation of the total action can be seen to be:
\bea
\delta S_{total} = -\frac{k}{2\pi} \int d^2x \,  \,[ (\tilde \kappa - \tilde \kappa_0) \, \delta f^{(-1)} + 4 \, \alpha^2 \,  (\tilde \omega - \tilde \omega_0) \, \delta g^{(-2)}]
\eea
where $\tilde \kappa_0$ and $\tilde \omega_0$ are some real numbers. Again we have several ways to satisfy $\delta S = 0$: 
\begin{enumerate}
\item $\delta f^{(-1)} = 0$ and $\delta g^{(-2)} = 0$. 

This is the Dirichlet condition again and leads to $W_3$ as the asymptotic symmetry algebra \cite{deBoer:2014fra}.
\item $\tilde \kappa = \tilde \kappa_0$ ($\tilde \omega = \tilde \omega_0$ ) and $\delta g^{(-2)} = 0$ ($\delta f^{(-1)} = 0$).

These are the other mixed boundary conditions -- we will not consider them further here.
\item $\tilde \kappa = \tilde \kappa_0$ and $\tilde \omega = \tilde \omega_0$ are the free boundary conditions we will consider below.
\end{enumerate}
\subsection{Solutions of $W_3$ Ward identities} 

The $W_3$ Ward identities (\ref{w3ward1}, \ref{w3ward2}) are also expected to be integrable (just as the Virasoro one in section 2 was) and general solutions can be written down by appropriate reparametrization of $f^{(-1)}$ and $g^{(-2)}$. 

However, we restrict to the case of $\tilde \kappa = \tilde \kappa_0$ and $\tilde \omega = \tilde \omega_0$ for constant $\tilde \kappa_0$ and $\tilde \omega_0$. This allows for classical solutions with fluctuating $f^{(-1)}$ and $g^{(-2)}$. Let us solve the $W_3$ Ward identities in this case. These read
\bea
\partial_-^3 f^{(-1)} + 24 \, \alpha^2 \, \tilde \omega_0 \, \partial_- g^{(-2)} - 4 \, \tilde \kappa_0 \, \partial_-f^{(-1)} &=& 0 \cr
\partial_-^5 g^{(-2)} - 20 \, \tilde \kappa_0 \, \partial_-^3 g^{(-2)} + 64 \, \tilde \kappa_0^2 \, \partial_-g^{(-2)} - 72 \, \tilde \omega_0 \, \partial_- f^{(-1)} &=& 0
\eea
There are two distinct cases: $\tilde \omega_0 = 0$ and $\tilde \omega_0 \ne 0$. When $\tilde \omega_0 = 0$ there are further two distinct cases:
\begin{enumerate}
\item $\tilde \omega_0 = 0$ and $\tilde \kappa_0 = 0$ gives:
\bea
\label{zerozero}
f^{(-1)} &=& f_{-1} (x^+) + x^- \, f_0 (x^+) + (x^-)^2  f_1 (x^+), \\
g^{(-2)} &=& g_{-2} (x^+) + x^- \, g_{-1}(x^+)+  (x^-)^2 \, g_0 (x^+) + (x^-)^3 \, g_1 (x^+) + (x^-)^4 \, g_2 (x^+) \nonumber 
\eea
This solution is suitable for non-compact $x^+$ and $x^-$ (such as the boundary of Poincare or Euclidean $AdS_3$).
\item $\tilde \omega_0 = 0$ and $\tilde \kappa_0 \ne 0$: 
\bea
\label{generalaps}
f^{(-1)} &=& f_\kappa(x^+) + g_\kappa (x^+) \, e^{2 \sqrt{\tilde \kappa_0} \, x^-} + \bar g_\kappa (x^+) \, e^{-2 \sqrt{\tilde \kappa_0}x^-}, \\
g^{(-2)} &=& f_\omega (x^+) + g_\omega (x^+) \, e^{2 \sqrt{\tilde \kappa_0} \, x^-} \!\!  + \bar g_\omega (x^+) \, e^{-2 \sqrt{\tilde \kappa_0} \, x^-} \!\!  + h_\omega (x^+) \, e^{4 \sqrt{\tilde \kappa_0} \, x^-} \!\! + \bar h_\omega (x^+) \, e^{-4 \sqrt{\tilde \kappa_0} \, x^-} \nonumber 
\eea
Again any positive value for $\tilde \kappa_0$ is suitable for non-compact boundary coordinates. Among the negative values $\tilde \kappa_0 = - \frac{1}{4}$ (times square of any integer) is suitable for compact boundary coordinates (such as the boundary of global $AdS_3$). 
\item When $\tilde \kappa_0 \ne 0$ and $\tilde \omega_0 \ne 0$, again the Ward identities can be solved. The general solutions involve eight arbitrary functions of $x^+$ (just as in the cases with $\tilde \omega_0 = 0$). Here we will only consider the special case where the solutions do not depend on $x^-$:
\bea
\label{generalcss}
f^{(-1)} = f(x^+), ~~~ g^{(-2)} = g(x^+).
\eea
This case is the analogue of \cite{Compere:2013bya} in the higher spin context.
\end{enumerate}
Next we analyse these cases one by one and find the asymptotic symmetries.
\section{Asymptotic symmetries, charges and Poisson brackets}
To find the asymptotic symmetries to which we can associate charges one needs to look for the residual gauge transformations of the solutions of interest. Just as in the $sl(2, {\mathbb R})$ case we can look at the residual gauge transformations of $a$ and $\tilde a$ and translate the corresponding gauge parameters $\lambda$ and $\tilde \lambda$ using $\Lambda = b^{-1} \lambda \, b$ and $\tilde \Lambda = b \, \tilde \lambda \, b^{-1}$. After finding these one can compute the corresponding charges.

A method for computing the charges corresponding to residual gauge transformations is provided by the Barnich {\it et} {\it al} \cite{Barnich:2001jy, Barnich:2007bf}. Using their method one can show that  the change in the charge $\mathrlap{\slash}\delta Q$ along the space of solutions of one copy of the Chern-Simons theory to be:
\bea
\mathrlap{\slash}\delta Q_\Lambda &=& - \tfrac{k}{2\pi} \int_0^{2\pi} d\phi ~ tr [\Lambda \, \delta A_\phi].
\eea
where $\Lambda$ is the gauge transformation parameter. We will see that these charges are integrable for all the residual gauge transformations considered below. 

Now, demanding that the charge (corresponding to a given residual gauge transformation) generates the right variations of the functions parametrizing the solutions via
\bea
\delta_{\Lambda} f(x) = \{ Q_\Lambda, f(x) \}, 
\eea
allows one to read out the Poisson brackets between those functions.

We are now ready to carry out this exercise for left sector and each of the cases (\ref{zerozero} - \ref{generalcss}) in the right sector one by one.
\subsection{The left sector symmetry algebra}
The left sector is common for all of the cases we consider in this paper. The corresponding  $\mathrlap{\slash}\delta Q$ is 
\bea
\mathrlap{\slash}\delta Q_\Lambda &=& - \tfrac{k}{2\pi} \int_0^{2\pi} d\phi \, (\lambda \, \delta\kappa-4\, \alpha^2 \, \eta \,  \delta \omega)
\eea
This when integrated between ($\kappa = 0$, $\omega =0$) and generic ($\kappa$, $\omega$) gives 
\bea
Q_{(\lambda, \eta)} = - \tfrac{k}{2\pi} \int_0^{2\pi} d\phi ~ [ \lambda \, \kappa - 4 \, \alpha^2 \, \eta \, \omega ]
\eea
This charge generates the variations (\ref{deltakappa}, \ref{deltaomega}) provided we take the Poisson brackets amongst $\kappa$ and $\omega$ to be:
\bea
-\tfrac{k}{2\pi}\left\{ \kappa ({x^+}),\kappa (\tilde x^+) \right\}&=& - \kappa'({x^+}) \, \delta (x^+ - \tilde x^+) - 2 \, \kappa \, ( x^+) \, \delta'({x^+}-\tilde x^+)+ \tfrac{1}{2} \, \delta'''({x^+}- \tilde x^+),\cr
-\tfrac{k}{2\pi}\left\{ \kappa({x^+}),\omega(\tilde x^+) \right\}&&= -2 \, \omega'({x^+}) \, \delta (x^+ - \tilde x^+) - 3  \,\omega (x^+) \, \delta'({x^+}- \tilde x^+),\cr
-\tfrac{2 k\alpha^2}{\pi} \, \left\{ \omega({x^+}),\omega(\tilde x^+) \right\}&=&\frac{8}{3} [ \kappa^2 (x^+) \, \delta' (x^+ - \tilde x^+) + \kappa (x^+) \, \kappa'(x^+) \delta (x^+ - \tilde x^+)] \cr 
&&
-\frac{1}{6} [5 \kappa (x^+) \delta''' (x^+ - \tilde x^+) + \kappa'''(x^+) \delta (x^+ - \tilde x^+)] \cr
&& - \frac{1}{4} [ 3 \kappa'' (x^+) \delta' (x^+ - \tilde x^+) + 5 \kappa'(x^+) \delta'' (x^+ - \tilde x^+)] + \frac{1}{24} \delta^{(5)} (x^+ - \tilde x^+) \nonumber \\
\eea
These brackets were computed by Campoleoni {\it et} {\it al}. \cite{Campoleoni:2010zq}. To compare with their answers one has to take $\kappa \rightarrow - \tfrac{2\pi}{k} \kappa$, $\omega \rightarrow \tfrac{\pi}{2k \alpha^2} \omega$, $\alpha^2 \rightarrow -\sigma$ in the expressions here. 

Next we turn to computing the charges and Poisson brackets on the right sector for all the cases of interest.

\subsection{$\tilde \kappa_0 = 0$ and $\tilde \omega_0 = 0$}
In this case the residual gauge transformation parameters are 
\bea
\tilde\lambda^{(-1)} &= &\lambda_{-1} (x^+) + x^- \,  \lambda_{0} (x^+)  + (x^-)^2 \, \lambda_1 (x^+)  \cr
\tilde \eta^{(-2)} &=& \eta_{-2} (x^+)  + x^- \, \eta_{-1} (x^+) + (x^-)^2 \, \eta_0 (x^+) + (x^-)^3 \, \eta_1 (x^+)  +  (x^-)^4 \, \eta_2 (x^+)
\eea
The corresponding action on the fields gives
\bea
\delta f_0 &=& \lambda_0'+ 2 \, (f_{-1} \, \lambda_1 -  \lambda_{-1} \, f_1) - 2 \, \alpha^2 \, (\eta_{-1} \, g_1 - \eta_1 \, g_{-1}) - 16  \, \alpha^2 \, (\eta_2 \, g_{-2} - \eta_{-2} \, g_2) \cr
\delta f_1 &=& \lambda_1' + (\lambda_1 \, f_0 - \lambda_0 \, f_1) - 2 \, \alpha^2 \,(\eta_0 \, g_1 - \eta_1 \, g_0) - 4 \, \alpha^2 \, (\eta_2 \, g_{-1} - \eta_{-1} \, g_2) \cr
\delta f_{-1} &=& \lambda_{-1}'+ (\lambda_0 \, f_{-1} - \lambda_{-1} \, f_0) - 2 \, \alpha^2 \, (\eta_{-1} \, g_0 - \eta_0 \, g_{-1}) - 4 \, \alpha^2 \, (\eta_1 \, g_{-2} - \eta_{-2} \, g_1) \cr
\delta g_0 &=& \eta_0' + 3 \, (\eta_1 \, f_{-1} - \eta_{-1} \, f_1) + 3 \, (\lambda_1 \, g_{-1} - \lambda_{-1} g_1) \cr
\delta g_1 &=& \eta_1' +  (\eta_1 \, f_0 - \lambda_0 \, g_1) + 2 \, (\lambda_1 \, g_0 - \eta_0 \, f_1) + 4 \, (\eta_2 \, f_{-1} - \lambda_{-1} \, g_2) \cr
\delta  g_{-1} &=& \eta_{-1}' + (\lambda_0 \, g_{-1} - \eta_{-1} \, f_0) + 2 \, (\eta_0 \, f_{-1} - \lambda_{-1} \, g_0) + 4 \, (\lambda_1 \, g_{-2} -  \eta_{-2} \, f_1) \cr
\delta g_2 &=& \eta_2' +  (\lambda_1 \, g_1 - \eta_1 \, f_1) + 2 \, (\eta_2 \, f_0 - \lambda_0 \, g_2) \cr
\delta g_{-2} &=& \eta_{-2}' +  (\eta_{-1} \, f_{-1} - \lambda_{-1} \, g_{-1}) + 2 \, (\lambda_0 \, g_{-2} - \eta_{-2} \, f_0)
\eea
 Defining
\bea
\label{poincarejl}
\{J^a, a = 1, \cdots, 8 \}  &=& \{f_{-1}, f_0, f_1, g_{-2}, g_{-1}, g_0, g_1, g_2\} \cr
\{\lambda^a, a = 1, \cdots, 8\} &=& \{ \lambda_{-1}, \lambda_0, \lambda_1, \eta_{-2}, \eta_{-1}, \eta_0, \eta_1, \eta_2 \} 
\eea
these expressions can also be written in a compact form:
\bea
\delta J^a = \partial_+ \lambda^a -  {f^a}_{bc} J^b \lambda^c 
\eea
where ${f^a}_{bc}$ are structure constants of our gauge algebra. The charge in this case is integrable and has the expression:
\bea
Q[\tilde \lambda] =  \tfrac{k}{4\pi} \int dx^+ \, \eta_{ab}  J^a \lambda^b
\eea
The Poisson brackets can be read out and we find:
\bea
\label{poincarecurrentalgebra}
\{ J^a (x^+), J^b(\tilde x^+) \} = {f^{ab}}_c \, J^c (x^+) \, \delta (x^+ - \tilde x^+) - \tfrac{k}{4\pi} \, \eta^{ab} \, \delta'(x^+ - \tilde x^+)
\eea
where we have redefined: $J^a \rightarrow -\tfrac{4\pi}{k} J^a$. This may be recognized as level $k$ Kac-Moody extension of the algebra used in defining the higher spin theory.
\subsection{$\tilde \kappa_0 = -\frac{1}{4}$ and $\tilde \omega_0 = 0$}
In this case the residual gauge transformation parameters are
\bea
\tilde\lambda^{(-1)} &=& \lambda_{f} (x^+)  + \lambda_g (x^+)  \, e^{i \, x^-} + \bar \lambda_{\bar g} (x^+)  \, e^{-i \, x^-} \cr
\tilde \eta^{(-2)} &=& \eta_f (x^+) + \eta_g (x^+) \, e^{i \, x^-} + \bar \eta_{\bar g} (x^+) \, e^{-i \, x^-} + \eta_h (x^+)  \, e^{2i\, x^-} + \bar \eta_{\bar h} (x^+)  \, e^{-2i \, x^-}
\eea
The symmetry transformations are:
\bea
\delta f_\kappa &=& \lambda_f'+ 2 \, i \, (\bar g_\kappa \, \lambda_g - \bar \lambda_{\bar g} \, g_\kappa) + 2 \, i \, \alpha^2 \, (\bar \eta_{\bar g} \, g_\omega - \eta_g \, \bar g_\omega) + 16 \, i \, \alpha^2 \,  (\eta_h \, \bar h_\omega - \bar \eta_{\bar h} \, h_\omega) \cr
\delta g_\kappa &=& \lambda_g' +i \, (\lambda_g \, f_\kappa - \lambda_f \, g_\kappa) + 2 \, i\, \alpha^2 \,  (\eta_f \, g_\omega - \eta_g \, f_\omega) + 4 \, i \, \alpha^2 \,  (\eta_h \, \bar g_\omega - \bar \eta_{\bar g} \, h_\omega) \cr
\delta \bar g_\kappa &=& \bar \lambda_{\bar g}'+i \, (\lambda_f \, \bar g_\kappa - \bar \lambda_{\bar g} \, f_\kappa) + 2 \, i \, \alpha^2 \,  (\bar \eta_{\bar g} \, f_\omega - \eta_f \, \bar g_\omega) + 4 \, i \, \alpha^2 \,  (\eta_g \, \bar h_\omega - \bar \eta_{\bar h} \, g_\omega) \cr
\delta f_\omega &=& \eta_f' + 3i \, (\eta_g \, \bar g_\kappa - \bar \eta_{\bar g} \, g_\kappa) + 3i \, (\lambda_g \, \bar g_\omega - \bar \lambda_{\bar g} g_\omega) \cr
\delta g_\omega &=& \eta_g' + i\, (\eta_g \, f_\kappa - \lambda_f \, g_\omega) + 2i \, (\lambda_g \, f_\omega - \eta_f \, g_\kappa) + 4i \, (\eta_h \, \bar g_\kappa - \bar \lambda_{\bar g} \, h_\omega) \cr
\delta \bar g_\omega &=& \bar \eta_{\bar g}' + i \, (\lambda_f \, \bar g_\omega - \bar \eta_{\bar g} \, f_\kappa) + 2i \, (\eta_f \, \bar g_\kappa - \bar \lambda_{\bar g} \, f_\omega) + 4i \, (\lambda_g \, \bar h_\omega - \bar \eta_{\bar h} \, g_\kappa) \cr
\delta h_\omega &=& \eta_h' + i (\lambda_g \, g_\omega - \eta_g \, g_\kappa) + 2i \, (\eta_h \, f_\kappa - \lambda_f\, h_\omega) \cr
\delta \bar h_\omega &=& \bar \eta_{\bar h}' + i \, (\bar \eta_{\bar g} \, \bar g_\kappa - \bar \lambda_{\bar g} \, \bar g_\omega) + 2i \, (\lambda_f \, \bar h_\omega - \bar \eta_{\bar h} \, f_\kappa)
\eea
Defining the currents $J^a$ and parameters $\lambda^a$ as 
\bea
\label{globaljl}
\{J^a, a = 1, \cdots, 8 \}  &=& \{ {\bar g}_\kappa, \, f_\kappa, \, g_\kappa, \,  {\bar h}_\omega, \,  {\bar g}_\omega, \,  f_\omega, \,  g_\omega, \,  h_\omega \} \cr
\{\lambda^a, a = 1, \cdots, 8\} &=& \{ \bar \lambda_{\bar g}, \, \lambda_f, \, \lambda_g,  \, \bar \eta_{\bar h},  \, \bar \eta_{\bar g},  \, \eta_f,  \, \eta_g,  \, \eta_h \} 
\eea
these expressions can also be written in a compact form:
\bea
\label{deltaja2}
\delta J^a &=& \partial_+ \lambda^a -i \,  \hat {f^a}_{bc} J^b \lambda^c 
\eea
where (some what surprisingly) $ \hat {f^a}_{bc}$ are obtained from the structure constants ${f^a}_{bc}$ by replacing $\alpha^2 \rightarrow - \alpha^2$. In this case the charge is:
\bea
Q[\lambda^a] = -\tfrac{k}{4\pi} \int_0^{2\pi} d\phi \, \hat \eta_{ab} \, \lambda^a J^b
\eea
where $\hat \eta_{ab}$ is the one obtained from $\eta_{ab}$ by replacing $\alpha^2$ by $- \alpha^2$.
The corresponding Poisson brackets are
\bea
\label{globalcurrentalgebra}
\{ J^a({x^+}),J^b(\tilde x^+) \} &=& i  {\hat f^{ab}}_{~~c} \,J^c(x^+) \, \delta(x^+-\tilde x^+) +\frac{k}{4\pi} 
\hat h^{ab} \delta'(x^+- \tilde x^+).
\eea
where again we have  redefined: $J^a \rightarrow \tfrac{4\pi}{k} J^a$. This again is a level-$k$ Kac-Moody algebra, but for the difference that it is obtained from the gauge algebra by $\alpha^2 \rightarrow - \alpha^2$ replacement.
\subsection{$\tilde \kappa_0 \ne 0$, $\tilde \omega_0 \ne 0$, $\partial_- f^{(-1)} = \partial_- g^{(-2)} = 0$}
In this case the residual gauge transformation parameters are 
\bea
\tilde\lambda^{(-1)} = \tilde \lambda (x^+), ~~~   \tilde \eta^{(-2)} = \tilde \eta (x^+).
\eea
Under these gauge transformations the fields transform as
\bea
\delta f^{(-1)} = \partial_+ \tilde \lambda, ~~ \delta g^{(-2)} = \partial_+ \tilde \eta.
\eea
Thus the residual gauge symmetries generate two commuting copies of $U(1)$ classically. Restricted to the $sl(2, {\mathbb R})$ sub-sector this case corresponds to \cite{Compere:2013bya}. The charge is
\bea
Q_{\tilde a}&=&  \tfrac{k}{2\pi} \int_0^{2\pi} d\phi\, 2 \, [\tilde \lambda \, (\tilde \kappa_0 \,  f - 6 \, \alpha^2 \, \tilde \omega_0 \, g) + \tilde \eta \,2 \, \alpha^2 \, (\tfrac{8}{3} \tilde \kappa_0^2 \,  g - 3 \, \tilde \omega_0 \, f)]
\eea
This leads to Poisson brackets:
\bea
\{ \tilde \kappa_0 \, f(x^+) - 6 \, \alpha^2 \,  \tilde \omega_0 \, g(x^+), \, f(\tilde x^+ )\} &=& -\tfrac{\pi}{k} \delta' (x^+ - {x^+}'), ~ \{ \tfrac{8}{3} \tilde \kappa_0^2 \, g(x^+) - 3 \, \tilde \omega_0 \, f(x^+), \, f(\tilde x^+) \} = 0 ~  \cr
\{ \tilde \kappa_0 \, f(x^+) - 6 \,\alpha^2 \, \tilde \omega_0 \, g(x^+), \, g(\tilde x^+ )\} &=& 0, ~
\{ \tfrac{8}{3} \tilde \kappa_0^2 \, g(x^+) - 3 \, \tilde \omega_0 \, f(x^+), \, g(\tilde x^+) \} = -\tfrac{\pi}{2k \alpha^2} \delta' (x^+ - \tilde x^+) \nonumber \\
\eea
These four relations are solved by the following three equations:
\bea
\{ f(x^+), \, f(\tilde x^+) \} = -\tfrac{\pi}{k} \tfrac{\tilde \kappa_0^2}{\Delta} \delta' (x^+ - \tilde x^+), &&
\{ g(x^+), \, g(\tilde x^+) \} = -\tfrac{\pi}{k} \tfrac{3\, \tilde \kappa_0}{16 \, \Delta \alpha^2} \delta' (x^+ - \tilde x^+), \cr
&&  \cr
\{ f(x^+), \, g(\tilde x^+) \} &=& -\tfrac{\pi}{k} \tfrac{9\, \tilde \omega_0}{8 \, \Delta} \delta' (x^+ - \tilde x^+)
\eea
where $\Delta = \tilde \kappa_0^3 - \tfrac{27}{4} \,\alpha^2 \, \tilde \omega_0^2$ which we have to assume not to vanish.\footnote{Taking linear combinations $f+ \chi \, g$ and $f- \chi \, g$ (for some constant $\chi$) as the currents one can decouple these two $u(1)$ Kac-Moody algebras.}
\section{Discussion}
In this paper, generalizing the results of \cite{Avery:2013dja}, \cite{Apolo:2014tua} we proposed boundary conditions for higher spin gauge theories in 3d in their first order formalism that are different from the usual Dirichlet boundary conditions.\footnote{It should be noted that the ansatz for the right sector gauge field (\ref{theansatz}) studied here also appeared recently in \cite{Compere:2013gja, deBoer:2014fra} where the authors were still interested in generalizations of dirichlet type boundary conditions.} The left sector is treated with the usual Dirichlet boundary conditions where as in the right sector we chose free boundary conditions. We restricted our attention to the spin-3 case for calculational convenience. The Dirichlet boundary conditions for general higher spin theory based on $sl(n, {\mathbb R})$ Chern-Simons was discussed in \cite{Campoleoni:2011hg} and for $hs[\lambda]$ case in \cite{Henneaux:2010xg}. One should be able to generalize our considerations to these other higher spin theories as well.

The boundary conditions considered here give one copy of $W_3$ and a copy of $sl(3, {\mathbb R})$ (or $su(1,2)$ or $u(1) \oplus u(1)$) Kac-Moody algebra. This matches with the symmetry algebra expected of the 2d chiral induced W-gravity with an appropriate field content. 

Let us emphasize that there appears to be a surprising difference between the asymptotic symmetry algebras of section 4.2 and section 4.3: namely the maximal finite subalgebra of (\ref{poincarecurrentalgebra}) is isomorphic to the gauge algebra of the higher spin theory where as that in (\ref{globalcurrentalgebra}) differs from the gauge algbra by $\alpha^2 \rightarrow -\alpha^2$ (this interchanges $sl(3, {\mathbb R})$ and $su(1,2)$). It will be interesting to understand the source of this possibility of getting a different real-form of the complexified gauge algebra out of our boundary conditions.

The Poisson brackets between $\kappa$ or $\omega$ of the left sector and any of the right sector currents vanish. Recall that in the $sl(2, {\mathbb R})$ case, motivated by how the asymptotic vector fields in the second order formalism \cite{Avery:2013dja} acted on the fields, we made (current dependent) redefinitions of the residual gauge parameters. Here too one can do such a redefinition. For instance, if we change variables
\bea
\lambda^a \rightarrow \lambda^a + \alpha_1 J^a \, \lambda + \alpha_2 \, {d^a}_{bc} J^b \, J^c \, \eta + \cdots 
\eea
where $\lambda^a$ are the parameters defined in (\ref{poincarejl}) and $\lambda$ and $\eta$ are the gauge parameters of the left sector, $d_{abc} \sim {\rm Tr} (T_a \{ T_b, T_c\})$, then one finds that 
\bea
\kappa \rightarrow \kappa + \# \, \eta_{ab} J^a J^b + \cdots, ~~ \omega \rightarrow \omega + \# \, d_{abc} J^a J^b J^c  + \cdots \, .
\eea
 The additional terms here may be recognized as the (classical analogues) of Sugawara constructions of spin-2 and spin-3 currents out of the Kac-Moody currents.

The general BTZ \cite{Banados:1992wn} type black hole solutions carrying higher spin charges (see \cite{Ammon:2012wc}) are not necessarily allowed classical solutions of our boundary conditions. For example, it can be seen that the boundary conditions considered in section 4.4 do allow such solutions, where as those in section 4.3 allow only some extremal ones.

It is of interest to understand the holographic duals of the higher spin theories with our boundary conditions better. For instance, how does one construct the action of the CIWG theories given the bulk theory and its boundary conditions. This question was addressed in the $n=2$ case by Banados {\it et} {\it al} \cite{Banados:2002ey}. For the case of $n=3$ we point out that the boundary conditions considered here can be seen to be consistent with the constraints imposed on the gauge connection in \cite{Ooguri:1991by} in their definition of CIWG as an $sl(3, {\mathbb R})$ gauged WZW model. It will be important to understand this connection better. In fact, Verlinde \cite{Verlinde:1989ua} anticipated that the CIWG theories could be defined through 3d gravity theories and our proposal can be considered as a realisation of that anticipation. 

We have used the first order formalism to do our computations. Our boundary conditions can be translated to the metric and the spin-3 fields in the second order language. There is a second order formalism of the 3d higher spin theories \cite{Campoleoni:2012hp, Fredenhagen:2014oua}. It will be interesting to work out the details in that formalism too.

Finally it will be interesting to see how to generalize these chiral boundary conditions to other contexts, such as other embeddings of gravity sector into the higher spin theory, supersymmetric theories {\it etc.}

\appendix
\section{$AdS_3$ gravity in first order formulation}

The $AdS_3$ gravity in the Hilbert-Palatini formulation can be recast as a gauge theory with action
\bea
S[A, \tilde A] = \frac{k}{4\pi} \int {\rm tr} (A\wedge A + \frac{2}{3} A \wedge A \wedge A) - \frac{k}{4\pi} \int {\rm tr} (\tilde A \wedge \tilde A + \frac{2}{3} \tilde A \wedge \tilde A \wedge \tilde A) 
\eea
up to boundary terms, where the gauge group is $SL(2, {\mathbb R})$. These are related to vielbein and spin connection through $A = \omega^a + \frac{1}{l} e^a$ and $\tilde A = \omega^a - \frac{1}{l} e^a$. The equations of motion are $F = dA + A \wedge A = 0$ and $\tilde F := d\tilde A + \tilde A \wedge \tilde A = 0$. We work with the following defining representation of the $sl(2,{\mathbb R})$ algebra. 
\begin{equation}
L_{-1} = \left(\begin{array}{rr} 0& -1\\ 0 & 0 \end{array}\right), ~~~ L_0 = \frac{1}{2} \left(\begin{array}{rr} 1& 0\\ 0 & -1 \end{array}\right), ~~ L_1=\left(\begin{array}{rr} 0& 0\\ 1 & 0 \end{array}\right), ~~ ,
\end{equation}
Satisfying $[L_m, L_n]= (m-n) L_{m+n}$. The metric defined by ${\rm Tr} (T_a, T_b) = \frac{1}{2} h_{ab}$ is 
\begin{equation}
h_{ab} = \left( \begin{array}{rrr} 0 ~ & 0 & -2 \\   0 ~ & 1 & 0 \\ -2 ~ & 0 & 0 \end{array} \right)
\end{equation}
It is known that the connections
\bea
A &=& b^{-1} \partial_r b \, dr + b^{-1} (L_1 - \kappa (x^+) \,  L_{-1}) \, b \, dx^+ \cr
\tilde A &=& b \, \partial_r b^{-1}  \, dr + b \,  ( \tilde \kappa (x^-) \, L_1-L_{-1} ) \, b^{-1} \, dx^-
\eea
represent all the solutions of $AdS_3$ gravity satisfying Brown-Henneaux (Dirichlet) boundary conditions (in FG coordinates) where $b = e^{L_0 \, \ln \frac{r}{l}}$. In fact, any solution of the Chern-Simons theory (locally) can be written as 
\bea
A = b^{-1} \partial_r b \, dr + b^{-1} \, a \, b, ~~ \tilde A = b \, \partial_r b^{-1}  \, dr + b \, \tilde a \, b^{-1}
\eea
where $a$ and $\tilde a$ are flat connections in two dimensions with coordinates $(x^+, x^-)$. The general solution can be written as $a = g^{-1} \, dg$ and $\tilde a = \tilde g \, d \tilde g^{-1}$ where $g$ and $\tilde g$ are $SL(2, {\mathbb R})$ group elements that depend on $(x^+, x^-)$.
We now present general solution to this flatness condition in a different parametrization that will be useful to us. Consider the most general $sl(2, {\mathbb R})$ 1-form on the boundary
\bea
a = (a^{(+)}_+ \, L_1 + a^{(-)}_+ \, L_{-1} + a^{(0)}_+ \, L_0) \, dx^+ + (a^{(+)}_- \, L_1 + a^{(-)}_- \, L_{-1} \, + a^{(0)}_- \, L_0) \, dx^-
\eea
Assuming that $a^{(+)}_+$ does not vanish, the flatness conditions imply:
\bea
a^{(0)}_- &=& \frac{1}{a^{(+)}_+} \left( a^{(+)}_- \, a^{(0)}_+ + \partial_- a^{(+)}_+ - \partial_+ a^{(+)}_- \right), ~~
a^{(-)}_- = \frac{1}{2a^{(+)}_+} \left(2 \, a^{(+)}_- \, a^{(-)}_+ + \partial_- a^{(0)}_+ - \partial_+a^{(0)}_-\right) \cr
&& ~~~~~~~ ~~~~~~~~~~ \frac{1}{2}\partial_+^3 f = \partial_-\kappa - 2 \, \kappa \, \partial_+ f - f \, \partial_+ \kappa
\eea
where $\kappa = a^{(+)}_+ \, a^{(-)}_+ -\frac{1}{4} (a^{(0)}_+)^2 - \frac{1}{2} \partial_+ a^{(0)}_+ + \frac{1}{2} a^{(0)}_+ \, \partial_+ \ln a^{(+)}_+ + \frac{1}{2} \partial_+^2 \ln a^{(+)}_+- \frac{1}{4} (\partial_+ \ln a^{(+)}_+)^2$ and $f = \frac{a^{(+)}_-}{a^{(+)}_+}$. Similarly if we consider the 1-form
\bea
\tilde a = (\tilde a^{(+)}_+ \, L_1 + \tilde a^{(-)}_+ \, L_{-1} + \tilde a^{(0)}_+ \, L_0) \, dx^+ + (\tilde a^{(+)}_- \, L_1 + \tilde a^{(-)}_- \, L_{-1} \, + \tilde a^{(0)}_- \, L_0) \, dx^-
\eea
Then, assuming now that $\tilde a^{(-)}_-$ does not vanish, the flatness conditions read
\bea
\tilde a^{(+)}_+ &=& \frac{1}{2\,\tilde a^{(-)}_-}(2 \, \tilde a^{(+)}_- \, \tilde a^{(-)}_+ + \partial_- \tilde a^{(0)}_+ - \partial_+\tilde a^{(0)}_-), ~~ \tilde a^{(0)}_+ = \frac{1}{\tilde a^{(-)}_-} \, (\tilde a^{(0)}_- \, \tilde a^{(-)}_+ + \partial_- \tilde a^{(-)}_+ - \partial_+\tilde a^{(-)}_-) \cr
&& ~~~~~~~~~~~~~~~~~~~~~ \frac{1}{2} \partial_-^3 \tilde f = \partial_+ \tilde \kappa -2 \, \tilde \kappa \, \partial_- \tilde f - \tilde f \, \partial_- \tilde \kappa 
\eea
where $\tilde f = \frac{\tilde a^{(-)}_+}{\tilde a^{(-)}_-}$ and $\tilde \kappa = \tilde a^{(-)}_- \, \tilde a^{(+)}_- - \frac{1}{4} (\tilde a^{(0)}_-)^2 + \frac{1}{2} \partial_-\tilde a^{(0)}_- - \frac{1}{2} \tilde a^{(0)}_- \, \partial_- \ln \tilde a^{(-)}_- + \frac{1}{2} \partial_-^2 \ln \tilde a^{(-)}_- - \frac{1}{4} (\partial_- \ln \tilde a^{(-)}_-)^2$.
The last equation is again the famous Virasoro Ward identity that can be solved explicitly as in section 2. Some special cases of the above formulae have appeared before, for instance, in \cite{Banados:2004nr}. 

\section{$sl(3,{\mathbb R})$ conventions}
We work with the following basis of $3\times 3$ matrices (see \cite{Campoleoni:2010zq}) for the fundamental representation of the gauge group used in the definition of the higher spin theory:
\bea
L_{-1} &=& \left( \begin{array}{crr} 0 & -2 & 0 \\ 0 & 0 & -2 \\ 0 & 0 & 0 \end{array} \right), ~~ L_{0} = \left( \begin{array}{ccr} 1 & ~ 0 & 0 \\ 0 ~ & 0 & 0 \\ 0 ~ & 0 & -1 \end{array} \right), ~~ L_{1} = \left( \begin{array}{crr} 0 & 0 & 0 \\ 1 & 0 & 0 \\ 0 & 1 & 0 \end{array} \right), ~~ W_{-2} = \alpha \, \left( \begin{array}{crr} 0 & 0 & 8 \\ 0 & 0 & 0 \\ 0 & 0 & 0 \end{array} \right), \cr
W_{-1} &=& \alpha \, \left( \begin{array}{crr} 0 & -2 & 0 \\ 0 & 0 & 2 \\ 0 & 0 & 0 \end{array} \right), ~~ W_0 = \alpha \, \tfrac{2}{3} \left( \begin{array}{crr} 1 & 0 & 0 \\ 0 & -2 & 0 \\ 0 & 0 & 1 \end{array} \right), ~~ W_1 = \alpha \, \left( \begin{array}{crr} 0 & 0 & 0 \\ 1 & 0 & 0 \\ 0 & -1 & 0 \end{array} \right), ~~ W_2 =\alpha \, \left( \begin{array}{crr} 0 & 0 & 0 \\ 0 & 0 & 0 \\ 2 & 0 & 0 \end{array} \right). \nonumber \\ 
\eea
For $\alpha^2 = -1$ these represent $su(1,2)$ algebra and for $\alpha^2 = 1$ they represent the $sl(3, {\mathbb R})$. We take the Killing metric as $\eta_{ab} = \tfrac{1}{2} \, {\rm Tr} (T_a T_b)$ where $T_a$ are the above matrices. The structure constants are $f_{abc} = \frac{1}{2}{\rm Tr} (T_a \, [T_b, T_c])$.

\bibliographystyle{utphys}
\providecommand{\href}[2]{#2}\begingroup\raggedright\endgroup

\end{document}